\documentclass[12pt,a4paper]{article}
\usepackage{graphicx}
\usepackage{amsmath}
\usepackage{amssymb}

\begin{document}

\title{Chaotic instantons in periodically perturbed double-well system}
\author{V.I. Kuvshinov\thanks{E-mail:V.Kuvshinov@sosny.bas-net.by}, A.V. Kuzmin\thanks{E-mail:avkuzmin@sosny.bas-net.by} and V.A.
Piatrou\thanks{E-mail:PiatrouVadzim@tut.by}\\
{\small Joint Institute for Power and Nuclear Research,} \\ {\small Krasina str. 99, Minsk,  220109, Belarus }}
\date{}
\maketitle

\begin{abstract}
Kicked double-well system is investigated both analytically and numerically. Phenomenological formula for ground quasienergy splitting is obtained  using resonances overlap criterion in the framework of chaotic instanton approach. Results of numerical calculations of quasienergy spectrum are in good agreement with the phenomenological formula.
\end{abstract}

\section*{Introduction}

Semiclassical properties of systems with mixed classical dynamics
is a reach rapidly developing field of research. One of
interesting results obtained in this direction is a chaos assisted
tunneling. It was shown that the structure of the classical phase
space of Hamiltonian systems can influence such purely quantum
processes as the tunneling~\cite{Lin:90, Bohigas:93}. It was
demonstrated in numerical simulations that existence of chaotic
motion region in the classical phase space of the system can
increase or decrease tunneling rate by several orders of
magnitude~\cite{Lin:90, Hanggi:91}. Typically one considers
tunneling between KAM-tori embedded into the "chaotic sea". The
region of chaotic motion affects tunneling rate because compared
to direct tunneling between tori it is easier for the system to
penetrate primarily into the chaotic region, to travel then along
some classically allowed path and to tunnel finally to another
KAM-torus~\cite{Utermann:94, Mouchet:01}.

Chaos assisted tunneling phenomenon as well as the closely related
coherent destruction of tunneling were experimentally observed in
a number of real physical systems. The observation of the chaos
assisted tunneling between whispering gallery-type modes of
microwave cavity having the form of the annular billiard was
reported in the Ref.~\cite{Dembowski:00}. The study of the
dynamical tunneling in the samples of cold cesium atoms placed in
an amplitude-modulated standing wave of light provided evidences
for chaos-assisted (three-state) tunneling as
well~\cite{Steck:02}. Recently, the coherent destruction of
tunneling was visualized in the system of two coupled periodically
curved optical waveguides~\cite{Valle:07}.

The most popular methods which are used to investigate the chaos
assisted tunneling are numerical methods based on Floquet
theory~\cite{Shirley:65}. Among other approaches to chaos-assisted
tunneling we would like to mention the path integral approach for
billiard systems~\cite{Frischat:98} and quantum mechanical
amplitudes in complex configuration space~\cite{Shudo:98}. In this
paper we will consider the original approach based on instanton
technique, which was proposed in \cite{KKS:02, KKS:03,
KK:05} and numerically tested in~\cite{Kuzmin:07,
Igarashi:06}.

Instanton is a universal term to describe quantum transition
between two topologically distinct vacuum states of quantum
system. In classical theory the system can not penetrate potential
or dynamical barrier, but in quantum theory such transitions may
occur due to tunneling effect. It is known that tunneling
processes can be described semiclassically using path integrals in
Euclidean (imaginary) time. In this case instantons are
soliton-like solutions of the Euclidean equations of motion with a
finite action. For example, in Euclidean Yang-Mills theory
distinct vacuum states are the states of different Chern-Symons
classes, and instanton solutions emerge due to topologically
nontrivial boundary conditions at infinity. In this paper we will
use a much simpler system to investigate the connection between
the tunneling, instantons and chaos, namely the kicked system with
double well potential. Our attention will be focused on the behavior of the quasienergy spectrum when perturbation of the kicked type is added to the system.

\section{Instantons in kicked double-well potential}

Hamiltonian of the particle in the double-well potential
can be written in the following form:
\begin{equation}\label{eq:H}
H_0 = \frac{p^2}{2 m} + a_0\, x^4 - a_2\, x^2,
\end{equation}
where $m$ - mass of the particle, $a_0, a_2$ - parameters of the
potential.

We consider the perturbation
\begin{equation}\label{eq:V}
V_{per} = \epsilon\, x \sum^{+ \infty}_{n = - \infty} \delta(t- n
T),
\end{equation}
where $\epsilon$ -  value of the perturbation, $T$ - period of the
perturbation, $t$ - time.

Full Hamiltonian of the system is the following:
\begin{equation}\label{SystemHamiltonian}
H = H_0 + V_{per}.
\end{equation}

Euclidean equations of motion of the particle in the double-well potential have a solution - instanton. In phase space of
nonperturbed system instanton solution lies on the separatrix.
Perturbation destroys the separatrix forming stochastic layer. In this
layer a number of chaotic instantons appears. Chaotic instanton
can be written in the following form:
\[x_{chaos} = x_{inst} + \epsilon\, \Delta x_{chaos}.\]
It is a solution of the Euclidean equations of motion. Here
$x_{chaos}$ and $x_{inst}$ - chaotic and nonperturbed instanton
solutions, respectively, $\Delta x_{chaos}$ - stochastic
correction.

It is convenient to work in the action-angle variables. Using standart technique \cite{Liberman:92} we obtain expressions for this variables in the following form:
\begin{eqnarray}\label{action}
J(E) = \frac{2 a^{3/2}_2 \sqrt{m}}{3 \pi a_0} \; \sqrt{1 + \sqrt{1 - 4
\frac{a_0}{a^2_2}\, E \,}} \left(L(\chi) - \sqrt{1 - 4
\frac{a_0}{a^2_2}\, E \,} \; K(\chi) \right),\\
 \Theta  = \frac{\pi}{2} \; F\left(\sqrt{\frac{2 a_0}{a_2}}\frac{x}{\sqrt{1 - \sqrt{1 - 4
\frac{a_0}{a^2_2}\, E \,}}}, \chi \right) \; K^{-1}(\chi),\label{angle}
\end{eqnarray}
where $J$ and $\Theta$ - action and angle variables, $E$ - energy of the particle, $K(\chi)$ and $L(\chi)$ - full elliptic integrals of the first and second kinds, respectively, $F(\phi,\chi)$ - elliptic integral of the first kind, where $\phi$ is integral argument, $\chi$ - modulus which is in the following way expresses through energy $E$
\[\chi = \frac{1 - \sqrt{1 - 4 \frac{a_0}{a^2_2}\, E
\,}}{1 + \sqrt{1 - 4 \frac{a_0}{a^2_2}\, E \,}}.\]

We will use in our analytical estimations expression for the action of the chaotic instanton. We assume this action is equal to the nonperturbed instanton action corresponding to the Euclidean energy which is less than one of the nonperturbed instanton solution. We expand expression (\ref{action}) in powers of the (Euclidean) energy difference from the separatrix $\xi=E_{sep}-E$, where $E$ - energy, $E_{sep}$ - energy on separatrix, and neglect terms higher than linear one. As a result we has the following linear expression
   \begin{equation}\label{eq:S}
   S[x_{chaos}(\tau, \xi)] = \pi J(E_{sep} - \xi) =  S[x_{inst}(\tau, 0)] - \alpha \,
\sqrt{\frac{m}{a_2}} \; \xi,
   \end{equation}
where \(S[x_{inst}(\tau, 0)] = 2 \sqrt{m} \, a^{3/2}_2 /(3 \, a_0)\) -
nonperturbed instanton action, $\alpha = (1 + 18 \ln 2)/6$ - numerical coefficient.

\begin{figure}[t]
\begin{center}
\includegraphics[angle = 270, width = 0.48\textwidth]{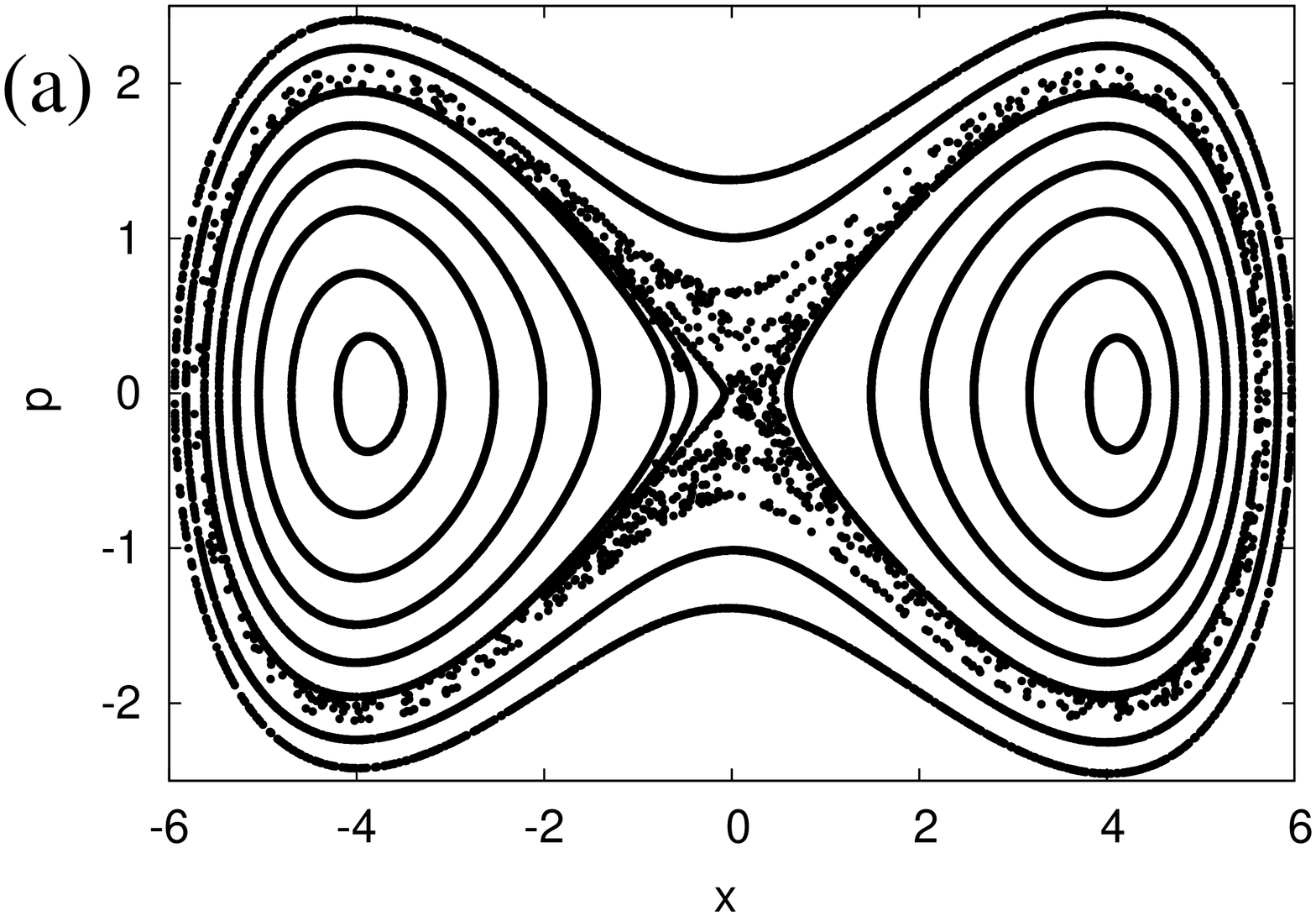}
\includegraphics[angle = 270, width = 0.48\textwidth]{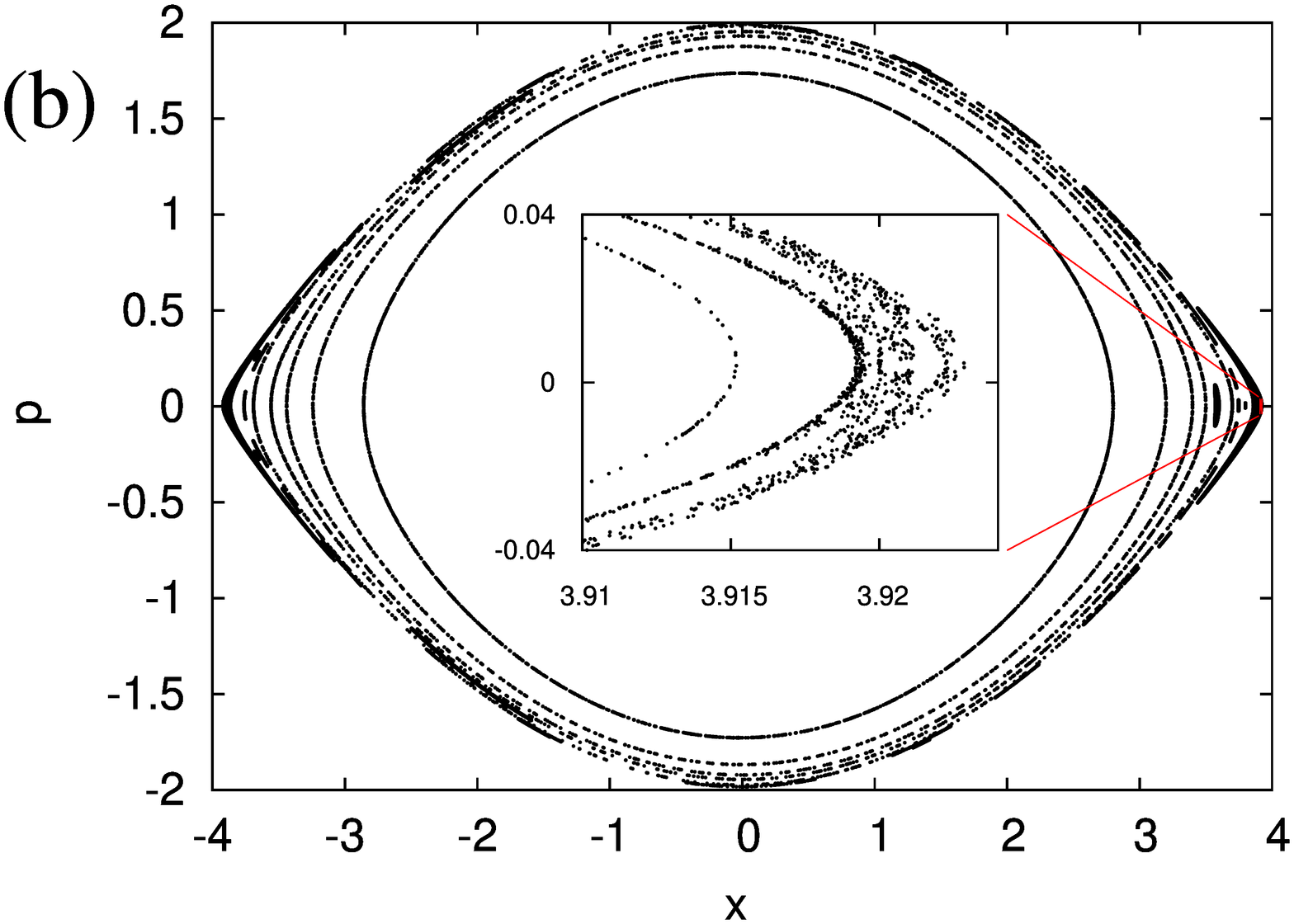}
\end{center}
   \caption{(a) Phase space portrait for the kicked double-well potential in the real time. (b) Phase space in the imaginary (Euclidean) time with inset where chaotic layer is showed. Chaotic instantons are placed in this layer. In both cases closed curves correspond to regular trajectories, scattered points to chaotic ones. The model parameters are $m =1, a_0 = 1/128,$ $a_2 = 1/4,$ $\epsilon = 0.01, \nu = 0.5$.}\label{fig:strob0p01}
\end{figure}

When stochastic layer in Euclidean phase space is formed under the action of the perturbation a number of chaotic
instanton solutions appear, see fig. \ref{fig:strob0p01}~(b). These solutions implement the saddle points of Euclidean action functional and thus give significant contribution to the tunneling amplitude. In the some sense their role is analogous to the role of the multiinstanton configurations in the case of the nonperturbed system. In the framework of the chaotic instanton approach the width of the stochastic layer determines the contribution of these chaotic solutions of Euclidean equations of motion to the tunneling amplitude. The stochastic layer width estimated using resonances' overlap criterion. For this purpose we should expand the expression for the perturbation~(\ref{eq:V}) in the series of exponents. At first we evaluate the coordinate  $x$ of the particle using expression~(\ref{angle}) in the following way:
\[x = \sqrt{\frac{a_2}{2 a_0}} \sqrt{1 - \sqrt{1 - 4 \frac{a_0}{a^2_2}\, E \,}} \: sn\left(\frac{2}{\pi} \, K(\chi) \Theta,
\chi\right)\]
and rewrite it as a series of  sines
\begin{equation}\label{eq:x}
x =  2\sqrt{2} \, \sqrt{\frac{m}{a_0}} \, \omega \sum^{\infty}_{n=1}\, \frac{a^{n-1/2}}{1-a^{2n-1}} sin[(2n-1)\omega \tau],
\end{equation}
where $a = exp[- \pi K(\chi')/K(\chi)]$ - parameter ($\chi' = 1 - \chi$), $\omega$ - the frequency of the particle oscillations.
Leaving only the first term in expression (\ref{eq:x}) and performing simplifications of this relation we obtain the following form for the coordinate
\begin{equation}\notag
 x = \frac{2\sqrt{2}}{\pi} \, \sqrt{\frac{a_2}{a_0}} \, sin[\omega(E) \tau].
\end{equation}
Using this relation and performing the Fourier transformation of the series of delta functions in the perturbation (\ref{eq:V}) one can obtain the following equations of motion in action-angle variables
\begin{eqnarray}\label{dotJ}
 \dot{J} = \frac{\epsilon \nu \sqrt{2}}{2 \pi^2} \, \sqrt{\frac{a_2}{a_0}} \, \sum^{\infty}_{n= - \infty} \left[e^{i (n \nu + \omega)\tau} + e^{i (n \nu - \omega) \tau}\right],\\
 \dot{\Theta} = \omega(J),\label{dotT}
\end{eqnarray}
where $\Theta$ - angle variable for the particle in the system, $\nu = 2\pi/T$ is a perturbation frequency.

To obtain width of stochastic layer we calculated the parameter of the resonances' overlap. It is equal to relation of the resonance width in the frequency scale ($\Delta \omega$) and the distance between resonances ($\delta \omega$). First one is estimated  as a frequency of oscillations of the resonance angle variable $\Psi = \Theta - n \nu \tau$. The result is obtained from the equations (\ref{dotJ}) and (\ref{dotT}) using standart technique (see chapter 1 and 5 in~\cite{Zaslavski:88}). The resonance width is the following
\begin{equation}\label{eq:DW}
\Delta \, \omega \sim \sqrt{\frac{\epsilon \, \nu \,\omega^2}{\Delta H}},
\end{equation}
where $\Delta H = E_{sep} - E$ - distance from the separatrix.

The distance between resonances is calculating using the expression for the resonance levels $\omega_n~=~n~\nu$. Thus one can obtain
\begin{equation}\label{eq:dW}
\delta \, \omega = \omega_{n+1} - \omega_n = \nu.
\end{equation}

Using two last expressions (\ref{eq:DW}) and (\ref{eq:dW}) the parameter of the resonances' overlap can be written in the following form
\begin{equation}\label{eq:K}
\overline{K} = \frac{\Delta \omega}{\delta \omega} \sim \left[\frac{\epsilon \, \omega^2}{\nu \, \Delta H} \right]^{1/2} \gtrsim \left[\frac{\epsilon \, \nu}{\Delta H} \right]^{1/2}.
\end{equation}
Overlap parameter is equal to unity on the boundary of the stochastic layer. Using equation (\ref{eq:K}) we can write the expression for the width of the stochastic layer in the following way
\begin{equation}\label{eq:dH}
\Delta H_s = E_{sep} - E_{bor} \approx \tilde{k} \,
\epsilon \, \nu,
\end{equation}
where $E_{bor}$  is the energy on the border between stochastic and regular regions,
$\tilde{k}$ - some numerical parameter which can not be obtained in the
framework of the criterion used.

The tunneling amplitude for the perturbed system is a sum of the amplitude in  the nonperturbed case and the amplitude of tunneling via chaotic instantons. The later can be evaluated by integration over action of the tunneling amplitude in nonperturbed
system. Using expression (\ref{eq:S}) this integral can be transformed to the integral over the energy difference from zero up to the width of the stochastic layer~(\ref{eq:dH}):
\begin{equation}\notag
 A_{chaos} = \alpha \sqrt{\frac{m}{a_2}}\,\tilde{N}\,\int^{\Delta H_s}_0 d\,\xi \int^{+ \infty}_{-\infty} d\,c_0  \sqrt{S[x_{chaos}(\tau, \xi)]} \, exp\,( - S[x_{chaos}(\tau, \xi)]),
\end{equation}
where $\tilde{N}$ is a normalize factor. To calculate contribution of chaotic instantons we use approximate expression for the chaotic instanton action (\ref{eq:S}) and listed above assumptions. Integration over $c_0$ gives the contribution of zero modes~\cite{Vainshtein:82}. As the result we get the following expression for the amplitude:
\begin{equation}\label{eq:A}
A  = A_{inst}+A_{chaos} \approx \tilde{N}\,\sqrt{S^{inst}} \, e^{-S^{inst}}  \, \Gamma \,
exp\left(\alpha \sqrt{\frac{m}{a_2}}\; \Delta H_s\right),
\end{equation}
where $A_{inst}$ is tunneling amplitude in the nonperturbed system, $\Gamma$ - a time of the tunneling which is put to infinity at the end. The last exponential factor in the expression~(\ref{eq:A}) is responsible for the tunneling enhancement in the perturbed system. In the nonperturbed case the width of the stochastic layer is equal to zero and the expression (\ref{eq:A}) coincides with the known expression describing the ordinary tunneling.

Substituting $\Delta H_s$ in formula (\ref{eq:A}) by the expression (\ref{eq:dH}) and expanding it in powers of the perturbation strength we can write phenomenological formula for the quasienergy splitting
\begin{equation}\label{eq:dE}
\Delta \eta(\epsilon, \nu) = 2\, \sqrt{\frac{6}{\pi}} \,
\sqrt{S^{inst}} \, e^{-S^{inst}}( 1 + k \; \epsilon \,\nu),
\end{equation}
where
$$k = \alpha \sqrt{\frac{m}{a_2}} \,\tilde{k}.$$
We fix this phenomenological parameter value using the results of
numerical simulations. For this purpose  we perform the linear fitting of the numerical data for the dependencies on the perturbation strength and take average value of the parameter over these dependencies. As the result we have the {\itshape single} numerical parameter~$k$ for our phenomenological formula explaining all these dependencies.

\section{Numerical calculations}

For the computational purposes it is convenient to choose as basis
vectors the eigenvectors of harmonic oscillator. In this
representation matrix elements of Hamiltonian (\ref{eq:H}) and the perturbation (\ref{eq:V}) are real and symmetric. They have the
following forms ($n \ge m$):
\begin{align*}
H^0_{m\, n} & = \delta_{m + 4 \; n} \;\frac{a_0 g^2}{4} \sqrt{(m + 1)(m + 2)(m + 3)(m +
4)} \\
 & + \delta_{m + 2 \; n} \;\frac{g}{2} \left(g\, a_0  (2 m + 3) - a'_2 \right) \sqrt{(m
+ 1)(m  + 2)}\\ + \delta_{m \;n}\; & \left[\hbar \omega (n + \frac12) + \frac g 2
\left(\frac32 \, g\, a_0 \, (2 m^2 + 2  m + 1) - a'_2 (2 m + 1) \right)  \right],\\
 V_{m\, n} & = \epsilon \; \delta_{m + 1 \; n} \;\sqrt{\frac g2} \; \sqrt{m + 1},
\end{align*}
where $g  = \hbar/m \omega$ and $a'_2 = a_2 + m \,\omega^2/2$,
$\hbar$ - Planck constant which we put equal to $1$, $\omega$ -
frequency of harmonic oscillator which is arbitrary, and so may be
adjusted to optimize the computation. We used the value $\omega =
0.2$ with parameters $m~=~1,$ $a_0~=~1/128,$ $a_2 = 1/4$. In numerical simulations size
of matrices was chosen to be equal to $200 \times 200$. Simulations with larger matrices give the same results.
\begin{figure}[h!]
\begin{center}
\includegraphics[angle = 270, width = 10cm]{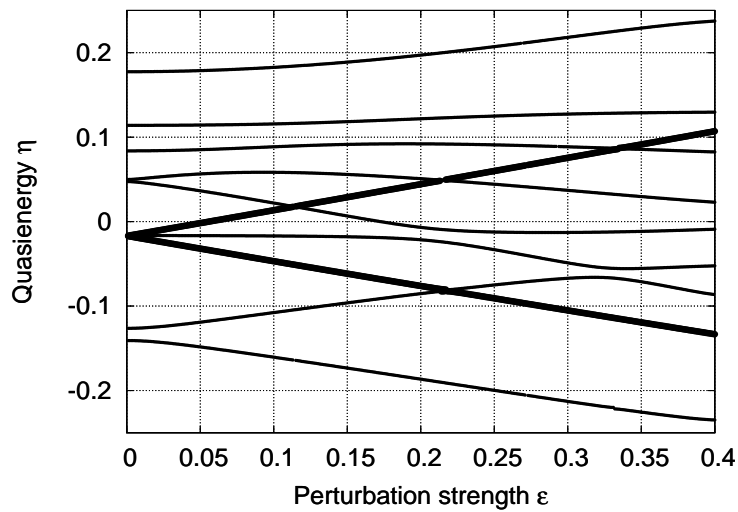}
\end{center}
\caption{Quasienergy spectrum for the ten levels with the lowest average energy. The model parameters are $m~=~1,$ $a_0~=~1/128,$ $a_2 = 1/4$ and $\nu = 0.5$. Thick lines - doublet with the minimal energy.}\label{fig:qeS}
\end{figure}

We calculate eigenvalues of the one period evolution operator $e^{-
i H T} e^{-i V}$ and obtain quasienergy levels which are related
with the evolution operator eigenvalues through the expression $\eta_k = i
\, \ln \lambda_k/T$. Then we get the ten levels with the lowest one period average energy which is calculated using the formula $\left<v_i\right|H_0 + V/T\left|v_i\right>$ ($\left|v_i\right>$ are the eigenvectors of the one period evolution operator). The dependence of quasienergies of this ten levels on the strength of the perturbation is shown in the figure~\ref{fig:qeS}. Quasienergies of two levels with the minimal average energy (thick lines in the figure~\ref{fig:qeS}) has a linear dependence on the strength of the perturbation in the considered region. They are strongly influenced by the perturbation while some of the quasienergy states are not.

\begin{figure}[t!]
\begin{center}
\includegraphics[angle = 270, width = 0.48\textwidth]{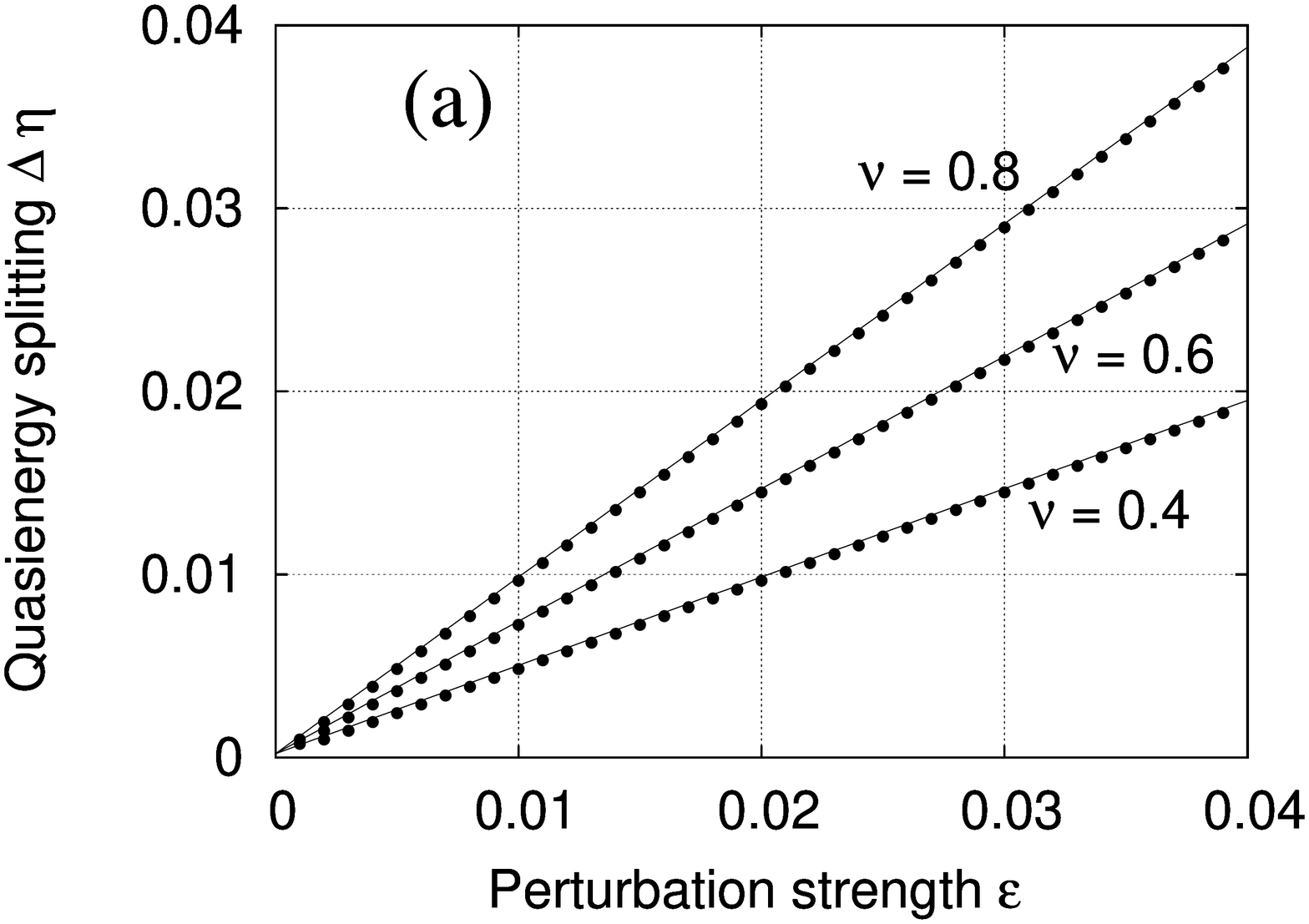}
\includegraphics[angle = 270, width = 0.48\textwidth]{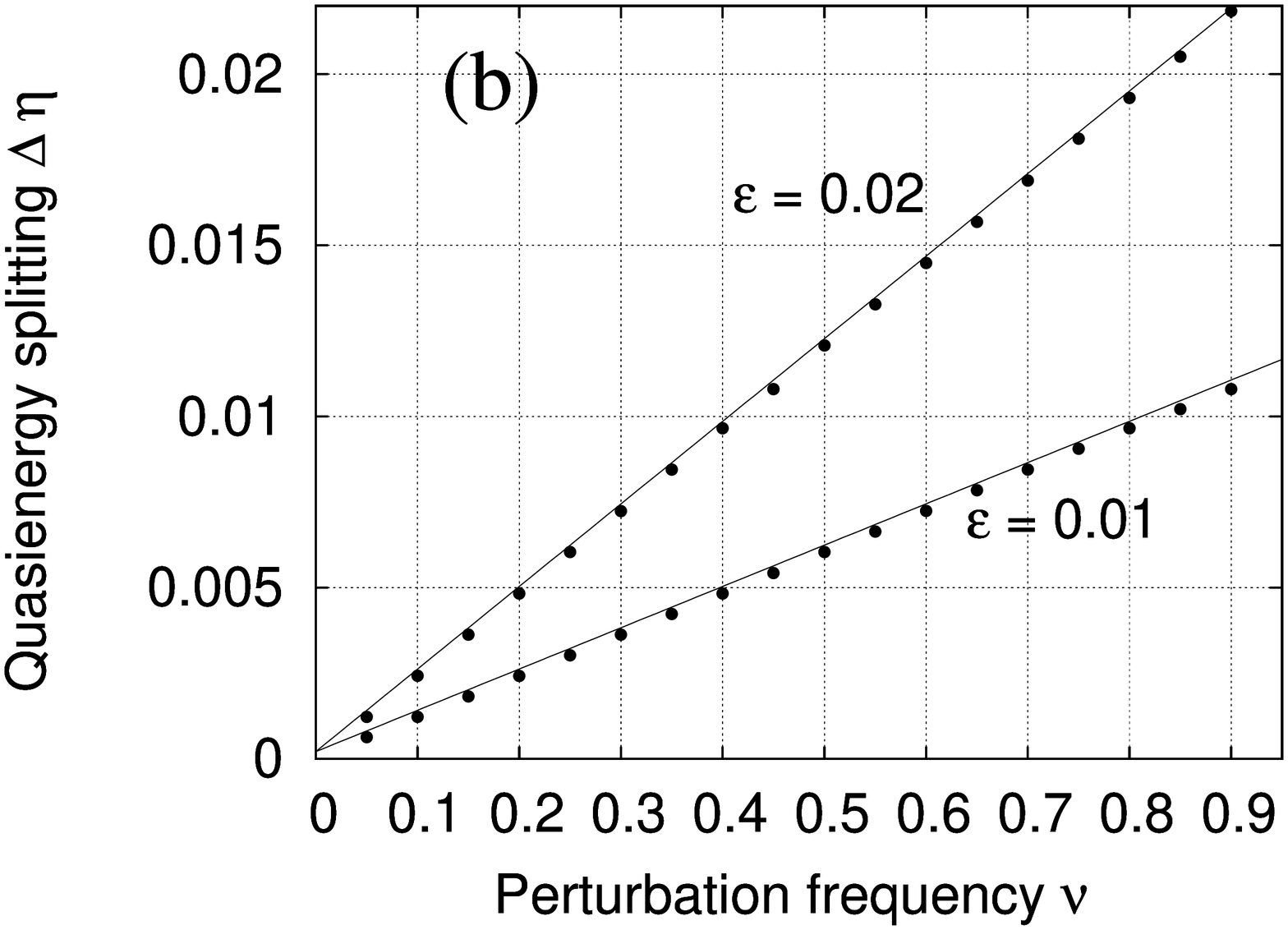}
\end{center}
\caption{Quasienergy splitting as a function of the strength~($a$) and
frequency~($b$) of the per\-tur\-ba\-tion. Lines - phenomenological
formula, points - numerical results. The model parameters are
$m~=~1,$ $a_0~=~1/128,$ $a_2 = 1/4$.}\label{fig:dE}
\end{figure}

Performed numerical calculations give the dependence of the quasienergy
splitting on the strength (fig.\ref{fig:dE}(a)) and the frequency (fig.\ref{fig:dE}(b)) of the perturbation. These dependencies are linear as it predicted by the obtained analytical formula (\ref{eq:dE}). Using least square technique we calculate the {\itshape single} numerical parameter $k$ which describes all numerical dependencies demonstrated in the figures \ref{fig:dE}(a) and \ref{fig:dE}(b). Relative error in determining of the parameter $k$ from numerical results is less than $0.1\%$. Analytical results are plotted in the figures~\ref{fig:dE}~(a)~and~(b) by lines. Numerical points lie close to this lines. The agreement between numerical simulations and analytical expression is good in the parametric region considered.

\section*{Conclusions}

Double-well system is investigated in presence of external kick perturbation. Analytic chaotic instanton approach is applied for
this system in order to obtain the phenomenological formula for the ground quasienergy splitting. The formula has the single numerical parameter which is determined from the numerical results. This formula describes ground quasienergy splitting as a function of strength and frequency of perturbation. It predicts linear dependence of the ground  quasienergy splitting on these parameters. Numerical results for the quasienergy splitting as a function of the perturbation frequency and strength demonstrate linear dependence as well. They are in a good agreement with the analytical formula. Analytical determination of the parameter $k$ at the moment seems to be impossible due to estimative character of the analytical method used. However, it is not a large problem since this parameter can be evaluated with the high accuracy using the single dependence of the ground quasienergy splitting for any particular system parameters' values. Parameter $k$ determined in such a way allows to describe correctly all other dependencies of the ground quasienergy splitting on the perturbation strength and frequency for any other values of the  system parameters. The only restriction is that the perturbation strength has to be sufficiently small (in the case considered $\epsilon \lesssim 0.1$) in order the chaotic instanton approach used to be valid.

\end{document}